\documentclass[24pt]{article}
\usepackage{amssymb}
\usepackage{latexsym}
\usepackage{amsmath}
\usepackage[cp1251]{inputenc}
\usepackage{graphicx}
\usepackage{color}

\title{High-density limit of quasi-two-dimensional dipolar Bose gas}

\author{Volodymyr~Pastukhov\footnote{e-mail:
volodyapastukhov@gmail.com}\\
{\small \textit{Ivan Franko
National University of Lviv, Department for Theoretical Physics}}\\
{\small \textit{12 Drahomanov St., Lviv, Ukraine }}}\date{}
\begin{document}
\maketitle
%\date{\today}
%\pacs{67.85.-d}
%\keywords{quasi-two-dimensional dipolar Bose system, Bogoliubov
%spectrum, Landau damping}
\begin{abstract}
We consider a simple model of the quasi-two-dimensional dipolar
Bose gas confined in the one-dimensional square well potential.
All dipoles are assumed to be oriented along the confining axis.
By means of hydrodynamic approach it is shown that the general
structure of the low-lying excitations can be analyzed exactly. We
demonstrate that the problem significantly simplifies in the
high-density limit for which the density profile in the confined
direction as well as the leading-order contribution to the
ground-state energy and spectrum of elementary excitations are
calculated. The low-temperature result for the damping rate of the
phonon mode is also presented.
\end{abstract}

\section{Introduction}
\label{sec1} \setcounter{equation}{0}

For more than decade the dipolar condensates of atoms with large
magnetic moments can be realized experimentally
\cite{Griesmaier,Lahaye,Lu,Aikawa}. Such experimental progress
stimulated extensive theoretical studies (see, for instance,
reviews \cite{Baranov,Lahaye_Menotti,Baranov_Dalmonte}). On the
other hand, it is well-known that three-dimensional Bose gas with
only dipole-dipole interaction is unstable, but presence of any
trapping potential that strongly confines system in the direction
of the external magnetic (electric) field stabilizes the system
\cite{Fischer,Koch,Wilson_Ronen}. The reason underlying this
property can be easily understood even on the mean-field level. In
this approximation, which accurately describes only dilute
systems, the stability condition is fully controlled by the sign
of zero-momentum Fourier transform of the effective two-body
interaction between particles. Although this potential
functionally depends on the one-dimensional density profile in the
confining direction, but at least weakly-interacting
quasi-two-dimensional dipolar Bose systems are always stable.
Another interesting feature of such objects is the existence of
maxon-roton behavior of the excitation spectrum \cite{Santos} even
in the simplest Bogoliubov's approximation. As the result the
Beliaev damping of these quasi-particles disappear in the vicinity
of the roton minimum \cite{Wilson_Natu}. Moreover, the presence of
roton-like behavior leads to modification of static
\cite{Klawunn_Recati} and dynamic \cite{Bisset_Baillie} structure
factors of the system, which can be measured using Bragg
spectroscopy. Recently, roton instability as well as formation of
stable droplets was observed experimentally \cite{Kadau} in the
dysprosium ($^{164}$Dy) dipolar Bose condensate. This
stabilization effect is possible due to the existence of quantum
fluctuations \cite{Lima_Pelster, Wachtler_Santos} and cannot be
understood on the mean-field level. The excitations with energy
close to double roton gap are unstable to the spontaneous decay
into two rotons \cite{Pitaevskii} leading to existence of the
termination point in the spectrum at very low temperatures that
was not observed in experiments with quasi-two-dimensional dipolar
Bose gases yet. The roton excitations were also theoretically
predicted in quasi-two-dimensional Bose condensate with
quadrupolar as well as contact interaction between particles
\cite{Lahrz_Lemeshko_Mathey}. The situation becomes more
complicated but no less interesting in the case when dipole
polarization forms non-zero angle with the confining direction.
This regime of the so-called tilted dipolar Bose condensate
includes the emergence of direction dependent superfluid
properties \cite{Ticknor_Wilson_Bohn}, non-typical vortex-vortex
interaction \cite{Mulkerin}, striped-phase formation
\cite{Macia_Hufnagl}, anomalous atom-number fluctuations
\cite{Baillie_Bisset_Ticknor} and anisotropic spectrum with
maxon-roton behavior \cite{Fedorov_Kurbakov,Baillie_Blakie}.
Finite temperature properties of quasi-two-dimensional dipolar
Bose systems were studied in Ref.~\cite{Ticknor_2012} within the
Hartree-Fock-Bogoliubov method, and in \cite{Pawlowski} using
classical field approximation. Fully two-dimensional dipolar Bose
gas with softened interaction in connection to the description of
polarized excitons was considered in Ref.~\cite{Kachintsev} with
the help of dielectric formalism. At finite temperatures the
presence of the roton minimum also changes properties of the
Landau damping both for phonon and maxon-roton parts of the
excitation spectrum \cite{Natu_Wilson}.

In the present paper, we consider quasi-two-dimensional dipolar
Bose system confined in a square well potential at low
temperatures. In particular, by means of the hydrodynamic approach
we develop perturbative treatment of the problem in terms of
inverse two-dimensional density of the system. Notwithstanding
that the experimental realization of Bose systems in square well
potential is complicated, our predictions may serve the starting
point to describe the harmonically confined systems near center of
the trap. Additionally we show that our findings for Bose systems
obtained within hydrodynamic approach are consistent with the
results of other formulations.

\section{Formulation}
\label{sec2} \setcounter{equation}{0}

 The considered model is described by the following Euclidian action
\begin{eqnarray}
S&=&\int dx \, \psi^*(x)\left(\frac{\partial}{\partial
\tau}+\frac{\hbar^2{\bf \nabla}^2}{2m}+\mu
\right)\psi(x)\nonumber\\
 &-&\frac{1}{2}\int dx\int dx' \Phi(x-x')
\psi^*(x)\psi^*(x')\psi(x')\psi(x),
\end{eqnarray}
where $x=(\tau, {\bf r})$, $\mu$ is chemical potential and the
following notation is used $\int dx=\int^{\beta}_0 d\tau \int_A
d{\bf r}_{\perp}\int^{a}_0 dz$. Here $\beta=1/T$ and $Aa$ are
inverse temperature and volume of the system, respectively.
Complex field $\psi(x)$ is $\beta$-periodic function of imaginary
time $\tau$ and the confining potential imposes boundary
conditions $\psi(x)|_{z=0}=\psi(x)|_{z=a}=0$. In transverse
direction we apply periodic boundary conditions with large area
$A$. This situation is approximately realized in pancake traps
with small frequencies in the plane perpendicular to the external
field direction. The second term of the action describes two-body
interaction between particles. We also introduce function
$\Phi(x-x')=\delta(\tau-\tau')\Phi({\bf r}-{\bf r}')$, where
dipole-dipole potential
\begin{eqnarray}\label{Phi_i}
\Phi({\bf
r})=\frac{g}{4\pi}\left\{\frac{1}{r^3}-\frac{3z^2}{r^5}\right\}.
\end{eqnarray}
Of course, in experimentally relevant cases the dipolar Bose gas
usually contains a short-range contact interaction as well, but
the presence of such a term in (\ref{Phi_i}) would not change our
further consideration.

Bearing in mind low-temperature description we can use one-mode
approximation and for convenience pass to density-phase variables
\cite{Popov,Shevchenko,Mora_Castin}
\begin{eqnarray}
\psi(x)=\chi_{0}(z)\sqrt{\rho(x_{\perp})}e^{i\varphi(x_{\perp})},
\end{eqnarray}
where $x_{\perp}=(\tau,{\bf r}_{\perp})$ and $\chi_0(z)$ is the
variational function with boundaries $\chi_{0}(0)=\chi_{0}(a)=0$
and normalization condition $\int^a_0dz|\chi_{0}(z)|^2=1$. The
multi-mode generalization of this approach required for
finite-temperature description of quasi-two-dimensional dipolar
Bose systems is straightforward, but necessarily leads to
considerable technical complications during calculations
\cite{Ticknor_2012}. After making use of Fourier transformation
\begin{eqnarray}
\rho(x_{\perp})=\rho+\frac{1}{\sqrt{\beta
A}}\sum_{K}e^{i(\omega_k\tau+{\bf k} {\bf r}_{\perp})}\rho_{K}, \\
\varphi(x_{\perp})=\frac{1}{\sqrt{\beta
A}}\sum_{K}e^{i(\omega_k\tau+{\bf k} {\bf r}_{\perp})}\varphi_{K},
\end{eqnarray}
where summations are carried out over $K=(\omega_k, {\bf k})$
($\omega_k$ is bosonic Matsubara frequency, ${\bf k}\neq 0$), we
can write down the effective action for quasi-two-dimensional
system
\begin{eqnarray}\label{S_eff}
S_{eff}=S_{0}+S_{G}+S_{int},
\end{eqnarray}
where constant mean-field part is
\begin{eqnarray}
S_{0}=\beta A\rho\left\{\mu-\frac{\hbar^2}{2m}\int^a_0 dz
\left|\frac{d\chi_0}{dz}\right|^2-\frac{1}{2}\rho\nu(0)\right\}.
\end{eqnarray}
The second term of Eq.~(\ref{S_eff}) is action of non-interacting
two-dimensional quasi-particles
\begin{eqnarray}\label{S_G}
S_{G}=-\frac{1}{2}\sum_{K}\left\{\omega_k\varphi_K\rho_{-K}
-\omega_k\varphi_{-K}\rho_{K}+\frac{\hbar^2k^2}{m}\rho\varphi_{K}\varphi_{-K}
\right.\nonumber\\
\left.
+\left[\frac{\hbar^2k^2}{4m\rho}+\nu(k)\right]\rho_{K}\rho_{-K}\right\},
\end{eqnarray}
and the last one takes into account collisions between them
\begin{eqnarray}\label{S_int}
S_{int}&=&\frac{1}{2\sqrt{\beta A}}\sum_{K,
Q}\frac{\hbar^2}{m}{\bf kq}\,
\rho_{-K-Q}\varphi_{K}\varphi_{Q}\nonumber\\
&+&\frac{1}{3!\sqrt{\beta
A}}\sum_{K+Q+P=0}\frac{\hbar^2}{8m\rho^2}(k^2+q^2+p^2)\rho_{K}\rho_{Q}\rho_{P}\nonumber\\
&-&\frac{1}{8\beta
A}\sum_{K,Q}\frac{\hbar^2}{2m\rho^3}(k^2+q^2)\rho_{K}\rho_{-K}\rho_{Q}\rho_{-Q}.
\end{eqnarray}
In fact action $S_{G}+S_{int}$ describes two-dimensional Bose
system and the Fourier transform of the effective two-particle
interaction $\nu(k)$ contains all information about confining
potential
\begin{eqnarray}\label{nu(k)}
\nu(k)=\int^a_0 dz
\int^a_0dz' \,|\chi_0(z)|^2\nu_k(z-z')|\chi_0(z')|^2,
\end{eqnarray}
where for sake of simplicity we have introduced two-dimensional
Fourier transform of the dipole-dipole interaction
\begin{eqnarray}
\nu_k(z) =\int_A d{\bf r}_{\perp}e^{i{\bf k}{\bf
r}_{\perp}}\Phi({\bf r})
=g\left\{\frac{2}{3}\delta(z)-\frac{1}{2}ke^{-k|z|}\right\}.
\end{eqnarray}
The last two terms of Eq.~(\ref{S_int}) are usually omitted in the
long-wavelength limit, but their contribution is significant at
finite temperatures \cite{Pastukhov}. Thermodynamic relation for
the grand canonical potential $-\partial \Omega/\partial \mu=N$
together with normalization condition on $\chi_0(z)$ fix the
quantity $\rho$ to be equilibrium area density $N/A$. Actually
this fact allows to provide further consideration of the
translation-invariant systems in canonical ensemble. But in our
case when the normalization condition has to be taken into account
the last step is to minimize the grand canonical potential of the
system with respect to function $\chi_0(z)$. These calculations
lead to Gross-Pitaevskii-type equation
\begin{eqnarray}\label{Eq_min}
\left\{-\frac{\hbar^2}{2m}\frac{d^2}{dz^2}+\int^a_0
dz'|\chi_0(z')|^2\phi(z'-z)\right\}\chi_0(z)=\mu\chi_0(z),
\end{eqnarray}
where we denote the effective one-dimensional potential in the
confined direction
\begin{eqnarray}\label{phi_z}
\phi(z)=\rho\nu_0(z)+\frac{1}{A}\sum_{{\bf k}\neq
0}\nu_k(z)\left[S_k-1\right].
\end{eqnarray}
Here
$S_k=\frac{1}{\beta}\sum_{\omega_k}\langle\rho_{K}\rho_{-K}\rangle/\rho$
is the static structure factor of two-dimensional bosons with
density $\rho$ and interacting via potential $\nu(k)$. The
function $\langle\rho_{K}\rho_{-K}\rangle$ is related to the
dynamic structure factor of the system and its poles determine the
spectrum of collective modes. Taking into account only first two
terms of the action (\ref{S_eff}) we correctly reproduce the
Bogoliubov theory with undamped spectrum
$E_{q}=\sqrt{\left(\frac{\hbar^2q^2}{2m}\right)^2+\frac{\hbar^2q^2}{m}\rho\nu(q)}$.
The further consideration is cumbersome, but hydrodynamic approach
allows us to build perturbation theory free of infrared
divergences \cite{Nepomnyashchy} which greatly simplifies the
calculations of the spectral properties of Bose systems. On the
other hand, it is not hard to show that general structure of the
low-lying excitations and consequently the long-range behavior of
the one-particle density matrix can be found exactly. This also
gives information about leading-order asymptote of the effective
one-dimensional potential $\phi(z)$ at large particle separations.

\subsection{Low-energy excitations}

For the first step we introduce the matrix correlation function
\begin{eqnarray}\label{matrix}
{\bf F}(K)=\left( \begin{array}{c c}
\langle\varphi_K\varphi_{-K}\rangle & \langle\rho_K\varphi_{-K}\rangle\\
\langle\varphi_K\rho_{-K}\rangle & \langle\rho_K\rho_{-K}\rangle\\
\end{array} \right),
\end{eqnarray}
that satisfies Dyson equation
\begin{eqnarray}
{\bf F}^{-1}(K)={\bf F}_0^{-1}(K)-{\bf \Pi}(K),
\end{eqnarray}
with zero-order approximation determined by the Gaussian part of
the action (\ref{S_G})
\begin{eqnarray}\label{matrix_0}
{\bf F}_0(K)= \left( \begin{array}{c c}
\frac{\hbar^2 k^2}{m}\rho & \omega_k\\
                -\omega_k & \frac{\hbar^2 k^2}{4 m\rho}+\nu(k)\\
\end{array} \right)^{-1}.
\end{eqnarray}
The elements $\Pi_{\varphi\varphi}(K)$, $\Pi_{\rho\rho}(K)$ of
matrix self-energy are even functions of frequency and
$\Pi_{\rho\varphi}(K)$ is odd function with additional constrain
$\Pi_{\rho\varphi}(-K)=\Pi_{\varphi\rho}(K)$. Moreover, taking
into account structure of the anharmonic terms of the action
(\ref{S_int}) and absence of infrared divergences in perturbation
theory it is readily seen that $\Pi_{\rho\varphi}(K)\propto \omega_k
k^2$ and consequently has no effect on the real part of phonon
mode. The leading-order contribution to the diagonal matrix
elements of correlation function (\ref{matrix}) can also be found
in the long-length limit. To derive the low-energy limit of
$\Pi_{\rho\rho}(K)$ we use the fact that differentiation of every
bare vertex function with respect to density while keeping
$\nu(k)$ fixed gives the vertex with one more zero-momentum
($K=0$) $\rho$ line. Of course, this conclusion holds for exact
vertices. The latter observation immediately leads to the identity
\begin{eqnarray}\label{susc}
\nu(0)-\Pi_{\rho\rho}(0)=\left(\frac{\partial \mu}{\partial
\rho}\right)_{T},
\end{eqnarray}
where $\left(\partial \mu/\partial \rho\right)_{T}$ is inverse
susceptibility of two-dimensional system with density $\rho$ and
fixed potential $\nu(k)$. Making use of gauge transformation
$\varphi(x_{\perp})\rightarrow \varphi(x_{\perp})+m{\bf v}{\bf
r}_{\perp}/\hbar$ and mentioning that the derivative
$\frac{i\hbar}{m}{\bf k} \frac{\partial}{\partial {\bf v}}$ adds
$\varphi(K)$ line with zero frequency and vanishingly small ${\bf
k}$ to every vertex, one can show that in low-energy limit
\begin{eqnarray}\label{Pi_varphi_s}
\frac{\hbar^2 k^2}{m}\rho-\Pi_{\varphi\varphi}(K\rightarrow
0)=\frac{\hbar^2 k^2}{m}\rho_s,
\end{eqnarray}
where $\rho_s$ coincides with two-dimensional superfluid density
of the system. Equations (\ref{susc}), (\ref{Pi_varphi_s}) connect
the low-energy spectrum of collective modes to the macroscopic
quantities of the system and will be tested below within
perturbation theory. The above analysis shows that equal-time
one-particle Green function of the system reveals non-diagonal
long-range behavior $\langle\psi(x)\psi^*(x')\rangle|_{\tau=\tau
'}\propto \chi_0(z)\chi^{*}_0(z')|{\bf r}_{\perp}-{\bf
r'}_{\perp}|^{-\alpha}$ with exponent $\alpha=mT/2\pi \hbar^2
\rho_s$ \cite{Popov_72,Petrov} which is typical for
Berezinskii-Kosterlitz-Thouless phase of two-dimensional
superfluid.

\section{High-density solution}

The main problem to be solved in our approach is finding the
solution of Eq.~(\ref{Eq_min}) and calculating the Fourier transform
of the two-dimensional potential (\ref{nu(k)}) that determines the
properties of the system. The situation is complicated by the fact
that $\phi(z)$ functionally depends on $\chi_0(z)$. Of course, in
the weak-coupling limit when dimensionless parameter
$\gamma=\frac{\rho g }{a}/\frac{\hbar^2}{ma^2}$ is small the
solution is trivial $\chi_{0}(z)=\sqrt{\frac{2}{a}}\sin(\pi z/a)$
with $\mu=\frac{\hbar^2\pi^2}{2m a^2}$, but physically interesting
is the inverse limit $\gamma\gg 1$ where excitation spectrum
exhibits maxon-roton behavior. The analysis simplifies greatly for
the high-density systems $\rho a^2\gg1$ not too close to roton
instability. More precisely the condition of validity of our
approximation can be formulated introducing dimensionless roton
gap $\delta=\Delta/\frac{\hbar^2}{2ma^2}$ and temperature
$t=T/\frac{\hbar^2}{2ma^2}$. Then the last term of the
one-dimensional potential (\ref{phi_z}) can be omitted in the
limits $\frac{1}{\rho a^2}\ln\delta^{-1}\ll 1$ and $\frac{1}{\rho
a^2}\frac{t}{\delta}\ll 1$ \cite{Boudjemaa_Shlyapnikov} at zero
and finite temperatures, respectively. These conditions are quite
general, i.e., independent on the specific perturbation scheme and
do not necessary require the interaction parameter $\gamma$ to be
small. Actually, for this high-density limit the mean-field
approximation becomes exact one determining thermodynamic
properties of the system, and function $\chi_0(z)$ is given by the
solution of the one-dimensional Gross-Pitaevskii equation
\begin{eqnarray}\label{Eq_min_simple}
-\frac{\hbar^2}{2m}\frac{d^2}{dz^2}\chi_0(z)+\frac{2}{3}\rho
g|\chi_0(z)|^2\chi_0(z)=\mu\chi_0(z),
\end{eqnarray}
with zero boundary conditions. For such simple geometry of
confining potential the solution reads
\begin{eqnarray}\label{solution}
\chi_0(z)=\frac{C(\kappa)}{\sqrt{a}}
\textrm{sn}(2K(\kappa)z/a,\kappa),
\end{eqnarray}
\begin{eqnarray*}\kappa=\sqrt{\frac{\gamma}{6}}C(\kappa)/K(\kappa), \ \
\mu=\frac{2\hbar^2}{ma^2}K^2(\kappa)(1+\kappa^2),
\end{eqnarray*}
where $\textrm{sn}(z,\kappa)$ is the Jacobi elliptic sine,
$C(\kappa)$ is the normalization constant and $K(\kappa)$ is the
elliptic integral of the first kind. Having found one-dimensional
mean-field density profile and chemical potential, we can
calculate the ground-state energy of the quasi-two-dimensional
dipolar Bose system $E_0/N=\frac{1}{\rho}\int^{\rho}_{0}\mu
d\rho$, or in our approximation
\begin{eqnarray}
E_0/N=\mu-\frac{\rho g}{3}\int^{a}_{0}dz |\chi_0(z)|^4.
\end{eqnarray}
The results of numerical evaluation for the dimensionless
ground-state energy per particle $E_0/N=\frac{\hbar^2}{2ma^2}e_0$
and chemical potential $\mu=\frac{\hbar^2}{2ma^2} \lambda$ of the
system are presented in Fig.~1.
\begin{figure}
\centerline{\includegraphics
[width=0.8\textwidth,clip,angle=-0]{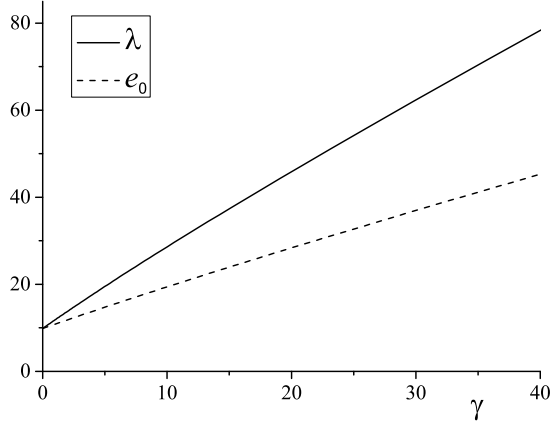}}
\caption{Dimensionless chemical potential $\lambda$ (solid line)
and ground-state energy per particle $e_0$ (dashed line).}
\end{figure}
It is worth noting that beyond mean-field calculation of the
thermodynamic properties of a quasi-two-dimensional dipolar Bose
system is tedious. First of all, one has to find the
next-to-leading order correction to the density profile in
confined direction. And secondly, these calculations require
knowledge of the quasi-two-dimensional dipole-dipole scattering
properties \cite{Wang,Macia_Mazzanti,Ticknor_2011}.

\subsection{Spectrum and damping rate}

We are in position now to calculate the function $\nu(k)$ and
study the spectral properties of the system. For reasons of
convenience we introduce function $f(\xi)=\nu(k)/\frac{g}{a}$,
where $\xi=ka$ is the dimensionless wave-vector. In two limiting
cases Fourier transform of the effective two-dimensional
interaction and as a result the Bogoliubov spectrum can be
calculated analytically. The first one is the above-mentioned
weak-coupling limit and the second one is known as the
Thomas–Fermi approximation which is applicable for $\gamma \gg 1$
where the function $\chi_0(z)$ tends to constant $1/\sqrt{a}$ on
the interval $[0,a]$. The explicit expressions for function
$f(\xi)$ for these two cases are the following
\begin{eqnarray*}
f_{0}(\xi)=\frac{1-e^{-\xi}}{\xi}-\frac{1}{2}\frac{\xi}{\xi^2+(2\pi)^2}
\left[\xi+3(1-e^{-\xi})-\frac{\xi^2-(2\pi)^2}{\xi^2+(2\pi)^2}(1-e^{-\xi})\right],
\end{eqnarray*}
\begin{eqnarray*}
f_{\textrm{TF}}(\xi)=-\frac{1}{3}+\frac{1-e^{-\xi}}{\xi}.
\end{eqnarray*}
The results of numerical calculations for the spectrum
$E_k=\frac{\hbar^2 }{2m a^2}\varepsilon(\xi)$,
$\varepsilon(\xi)=\xi \sqrt{\xi^2+4\gamma f(\xi)}$ of Bogoliubov
excitations are presented in Fig.~2.
\begin{figure}
\centerline{\includegraphics
[width=0.8\textwidth,clip,angle=-0]{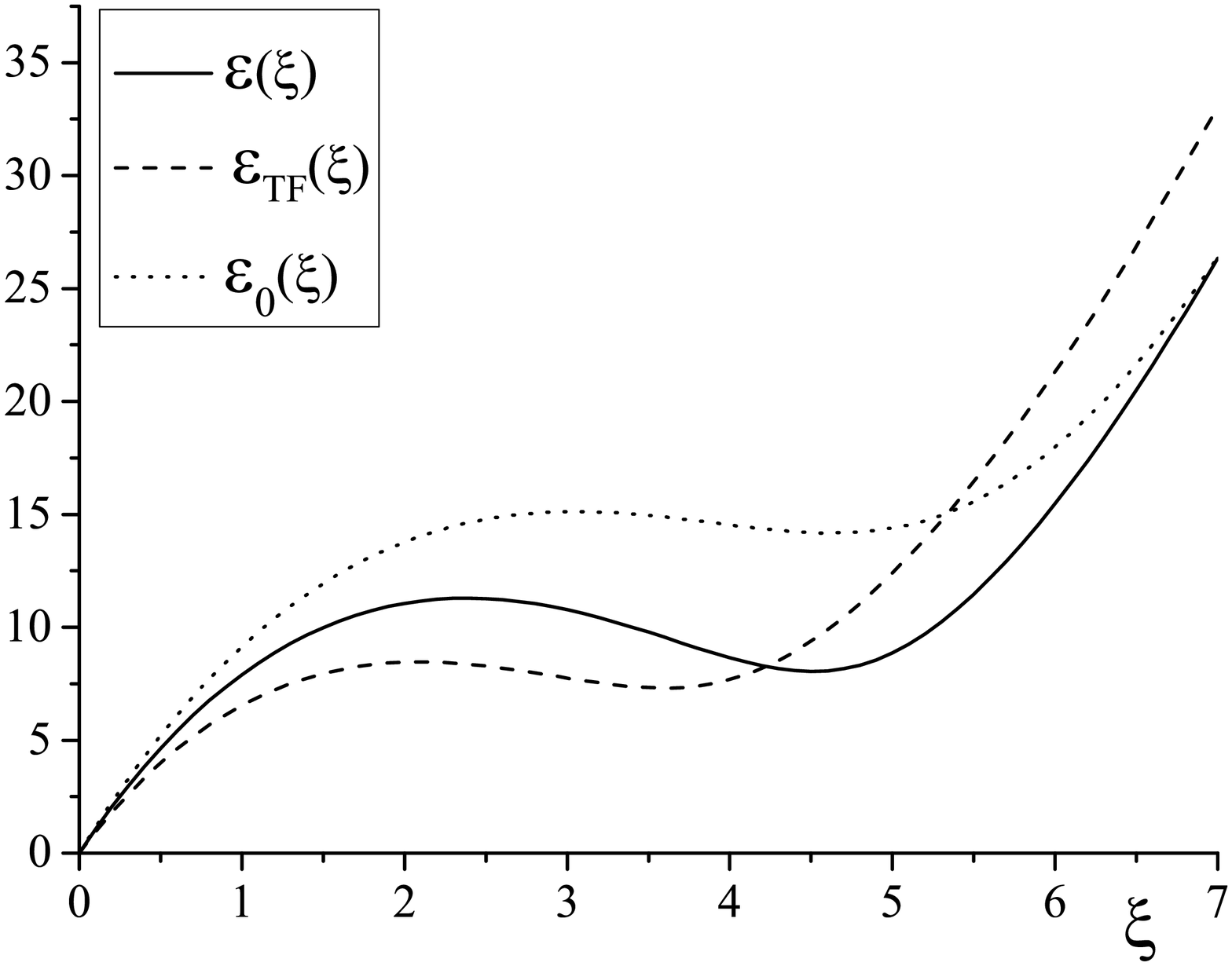}}
\caption{Bogoliubov spectrum in units of $\frac{h^2}{2ma^2}$ for
$\gamma=35$. Solid line presents exact calculation with solution
(\ref{solution}). Dotted and dashed lines are the results for
weak- and strong-coupling limits, respectively.}
\end{figure}
In the long-wavelength limit we have $E_k=\frac{\hbar
c}{a}\xi\left(1-\xi/4f(0)+\ldots\right)$. Here we introduce the
sound velocity $c=\sqrt{\rho\nu(0)/m}=\sqrt{\frac{\rho g}{m
a}}u(\gamma)$ where $u(\gamma)$ is dimensionless monotonically
decreasing function (see Fig.~3).
\begin{figure}
\centerline{\includegraphics
[width=0.8\textwidth,clip,angle=-0]{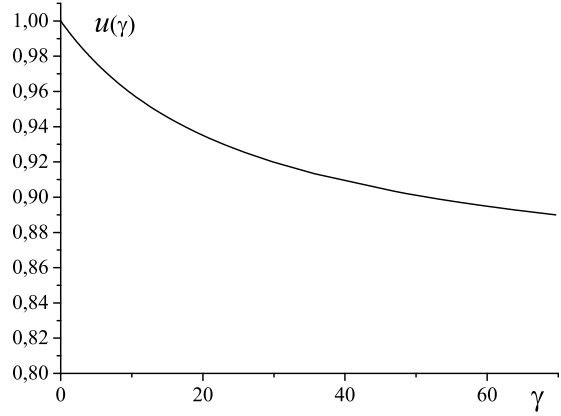}}
\caption{Dependence of the dimensionless sound velocity on the
 coupling parameter.}
\end{figure}
For the maximal and minimal values we obtained analytically
$u(0)=1$ and $u(\infty)=\sqrt{2/3}\simeq 0.816$, respectively.
Note that the second term of the low-energy spectrum is quadratic.
It is general feature of the quasi-two-dimensional dipolar Bose
systems that does not depend on the nature of confining potential.
Our numerical calculations show that at $\gamma_c\simeq 39.86$ the
roton instability occurs. In fact, this value of coupling
parameter determines the limits of applicability of our
approximation.

Finally, we have to make sure that damping of the spectrum is
small, i.e., elementary excitations are well-defined. The
calculation of the damping rate requires the knowledge of the
self-energy parts for which the explicit formulae on the one-loop
level are given in the Appendix. Although in our approximation the
real parts of self-energies do not affect on the properties of the
system, but to verify the identities (\ref{susc}),
(\ref{Pi_varphi_s}) we calculate the leading-order contribution to
the function $\Pi_{\varphi\varphi}(K)$ in the long-length limit
($K\rightarrow 0$)
\begin{eqnarray}
\Pi_{\varphi\varphi}(K)=\frac{\hbar^2k^2}{m}\frac{1}{2A}\sum_{{\bf
q}\neq 0}\frac{\hbar^2q^2}{m}\left[-\frac{\partial}{\partial
E_{q}}n(\beta E_{q})\right],
\end{eqnarray}
where $n(x)=(e^x-1)^{-1}$ is the Bose distribution function, and
it is easy to recognize the well-known Landau formula for the
normal density of the superfluid in two dimensions. In the same
manner one can show that the first-order calculations for
$\Pi_{\rho\rho}(K)$ lead to the result
\begin{eqnarray}
\Pi_{\rho \rho}(0)=\frac{1}{A}\sum_{{\bf q}\neq
0}\left[\frac{\hbar^2q^2\nu(q)}{2mE_q}\right]^2\left\{\frac{1}{2E_q}+\frac{1}{E_q}n(\beta
E_q)-\frac{\partial}{\partial E_{q}}n(\beta E_{q}) \right\},
\end{eqnarray}
which up to a sing coincides with the correction to the inverse
compressibility of the two-dimensional system. For the damping
rate of low-energy quasi-particles we have
\begin{eqnarray}\label{damp_rate}
\frac{\Gamma_k}{E_k}=\frac{\rho}{2mc^2}\Im\Pi_{\rho\rho}(E_k,k)-\frac{1}{\hbar
c k}\Re\Pi_{\varphi\rho}(E_k,k)+\frac{m}{2\hbar^2k^2\rho}
\Im\Pi_{\varphi\varphi}(E_k,k),
\end{eqnarray}
where $\Im\Pi_{\rho\rho}(\omega,k)$,
$\Im\Pi_{\varphi\varphi}(\omega,k)$ are imaginary parts and $\Re
\Pi_{\varphi\rho}(\omega,k)$ is real part of appropriate
self-energies after analytical continuation $i\omega_k\rightarrow
\omega+i0$.  Due to the sign of the second term in the
low-wavelength expansion of the Bogoliubov spectrum the Beliaev
damping of phonon mode is strongly exhausted \cite{Natu_Sarma}. At
finite temperatures the damping rate is fully controlled by the
so-called Landau mechanism of quasi-particle decay. So, in
Appendix we present the only terms of matrix self-energy
responsible for the Landau damping of low-lying excitations [see
Eqs.~(\ref{ImPi_vv})-(\ref{RePi_vr})]. After substitution in the
equation (\ref{damp_rate}) we obtain $\Gamma_k/E_k=i(t)/\rho a^2$,
where dimensionless damping rate reads
\begin{eqnarray}\label{i_tau}
i(t)=\frac{\gamma}{2\pi t f(0)}\int^{\infty}_0d \xi \, \xi^3
\frac{\Theta \left(|\varepsilon'(\xi)|/2\sqrt{\gamma
f(0)}-1\right)}{\sqrt{\varepsilon'^2(\xi)/4\gamma f(0)-1}}
\left[\frac{f(\xi)\xi}{\varepsilon(\xi)}+\frac{2f(0)}{\varepsilon'(\xi)}\right]^2\nonumber\\
\times n(\varepsilon(\xi)/t)[1+n(\varepsilon(\xi)/t)],
\end{eqnarray}
here $\Theta(x)$ is the Heaviside step function and we use
notation for the derivative of dimensionless spectrum
$\varepsilon'(\xi)$ with respect to $\xi$.
\begin{figure}
\centerline{\includegraphics
[width=0.8\textwidth,clip,angle=-0]{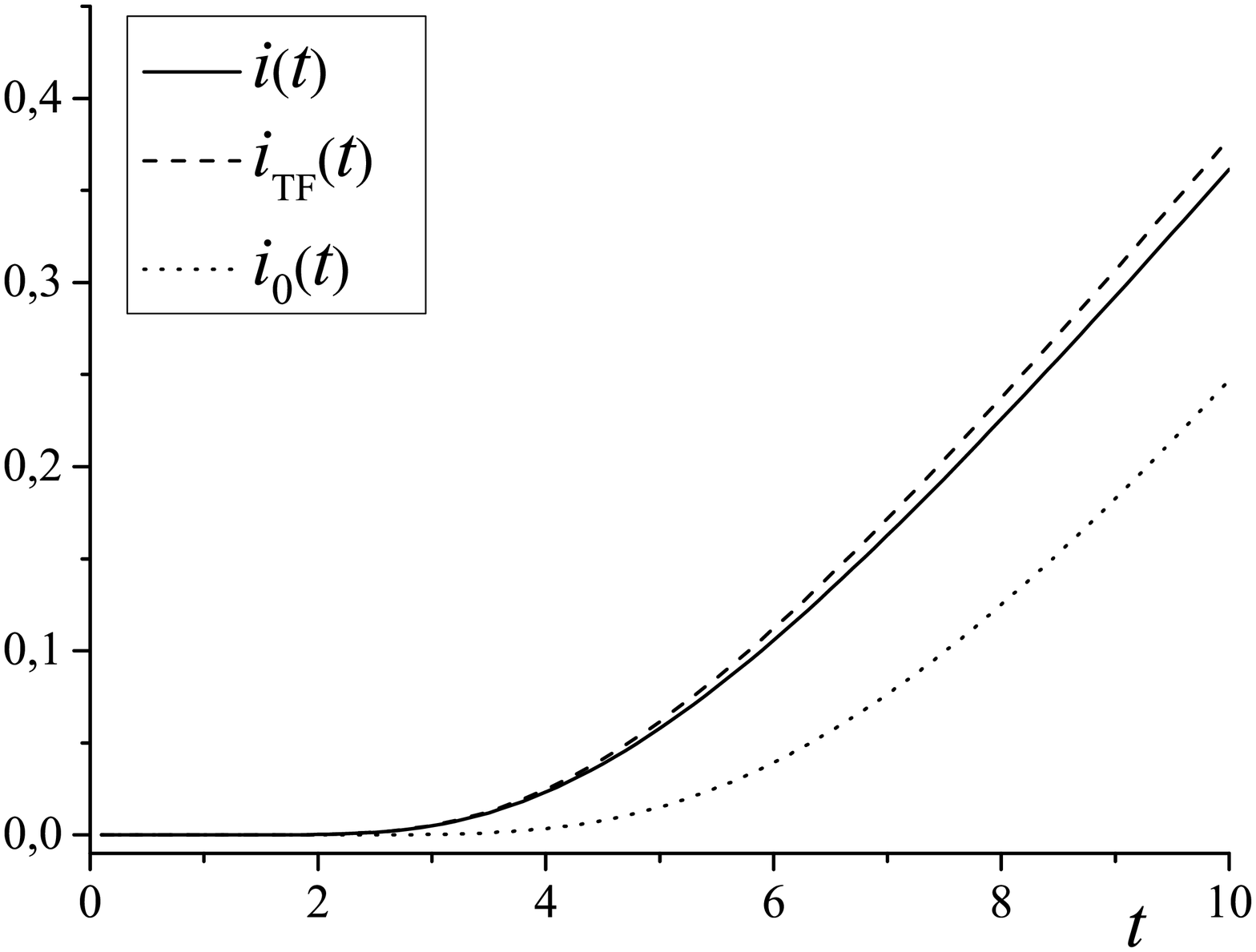}}
\caption{Temperature dependence of the damping rate of the phonon
mode calculated with exact (solid line), Thomas-Fermi (dashed
line) and weak-coupling (dotted line) density profiles at
$\gamma=35$.}
\end{figure}
This result is similar to that obtained with the help of other
approach \cite{Natu_Wilson}. If we set $f(\xi)=1$ it reproduces
the damping rate of the two-dimensional Bose gas with contact
interaction obtained with the help of the Hartree-Fock-Bogoliubov
approximation in Ref.~\cite{Chung_Bhattacherjee} and using
modified hydrodynamic approach in \cite{Pastukhov}. Our findings
prove the consistency of Popov's hydrodynamic description of Bose
systems at finite temperatures. As it is seen from formula
(\ref{i_tau}) the only contribution to the damping of low-energy
excitations comes from the quasi-particles with the velocity
$\partial E_q/\partial (\hbar q)$ greater than sound velocity of
the system \cite{Natu_Wilson}. This peculiarity of the damping
rate is confirmed by numerical calculations and illustrated in
Fig.~4. Similarly to Bose systems with contact interaction, at
high temperatures the curves clearly demonstrate linear behavior
on the reduced temperature.

\section{Conclusions}
\label{sec4}

In conclusion, we have studied properties of quasi-two-dimensional
dipolar Bose system confined in square well potential. At low
temperatures such a simple geometry of the external potential
allows to construct systematic perturbation theory in terms of the
inverse two-dimensional (area) density of the system. In the
leading order the mean-field equation that determines
one-dimensional density profile is solved exactly and the
effective potential of inter-particle interaction in two
dimensions is calculated. For the calculations of spectral
properties of the system we used the hydrodynamic approach. This
formulation is naturally-suited for two-dimensional Bose systems
and allows to analyze the low-energy spectrum of collective modes
exactly.

Our approximation leads to the Bogoliubov excitations with
maxon-roton behavior of the spectrum which we calculate
numerically using exact solution of the Gross-Pitaevskii equation
for the ground-state density profile. The same calculations were
also performed with approximate solutions of
Eq.~(\ref{Eq_min_simple}) for weak-coupling and Thomas-Fermi
limits, respectively. Although our findings clearly demonstrate
qualitative agreement in the behavior of the spectrum and damping
rate in these three cases, but quantitative differences between
curves (see Figs.~2,4) are noticeable.

\begin{center}
{\bf Acknowledgements}
\end{center}

We want to thank Dr.~Andrij~Rovenchak for useful comments on the
manuscript and Orest~Hryhorchak for help in numerical calculations. This work was partly supported by Project FF-30F (No.~0116U001539) from the
Ministry of Education and Science of Ukraine.

\section*{Appendix}

In this section we present the results of calculation for
self-energies to the first order of perturbation theory
\begin{eqnarray}
    \Pi_{\varphi\varphi}(K)=\frac{1}{\beta
        A}\sum_{Q}\frac{\hbar^4}{m^2}\left\{ ({\bf kq})^2\langle\varphi_Q\varphi_{-Q}\rangle\langle\rho_{K+Q}\rho_{-K-Q}\rangle \right.\nonumber\\
   \left. -{\bf kq}({\bf kq}+k^2)\langle\varphi_Q\rho_{-Q}\rangle\langle\varphi_{K+Q}\rho_{-K-Q}\rangle\right\},
\end{eqnarray}
\begin{eqnarray}
    \Pi_{\rho\rho}(K)=\frac{1}{2\beta
        A}\sum_{Q}\frac{\hbar^4}{m^2}\left\{(q^2+{\bf qk})^2\langle\varphi_Q\varphi_{-Q}\rangle\langle\varphi_{K+Q}\varphi_{-K-Q}\rangle  \right.\nonumber\\
    \left.+\frac{1}{16\rho^4}(k^2+q^2+{\bf kq})^2\langle\rho_Q\rho_{-Q}\rangle\langle\rho_{K+Q}\rho_{-K-Q}\rangle\right.\nonumber\\
    \left.+\frac{1}{2\rho^2}(k^2+q^2+{\bf kq})(q^2+{\bf qk})\langle\varphi_Q\rho_{-Q}\rangle\langle\varphi_{K+Q}\rho_{-K-Q}
    \rangle\right\}\nonumber\\
    -\frac{1}{4\beta A}\sum_{Q}\frac{\hbar^2}{m\rho^3}(k^2+q^2)\langle\rho_Q\rho_{-Q}\rangle,
\end{eqnarray}
\begin{eqnarray}
\Pi_{\varphi\rho}(K)=\frac{1}{\beta
    A}\sum_{Q}\frac{\hbar^4 {\bf kq}}{m^2}\left\{ (q^2+{\bf kq})\langle\varphi_Q\varphi_{-Q}\rangle\langle\rho_{K+Q}\varphi_{-K-Q}\rangle \right.\nonumber\\
\left. +\frac{1}{4\rho^2}(k^2+q^2+{\bf
kq})\langle\varphi_Q\rho_{-Q}\rangle\langle\rho_{K+Q}\rho_{-K-Q}\rangle\right\}.
\end{eqnarray}
where appropriate correlation functions should be taken neglecting
self-energy corrections, i.e., given by Eq.~(\ref{matrix_0}).
After performing the Matsubara frequency summations and passing to
the upper complex half-plane $i\omega_k\rightarrow \omega+i0$ we
obtain for self-energies
\begin{eqnarray}\label{ImPi_vv}
\Im \Pi_{\varphi\varphi}(\omega,k)=-\omega\frac{\pi}{A}\sum_{{\bf
q}\neq 0} \left[\frac{\hbar^2{\bf
kq}}{m}\right]^2\left[\frac{\partial}{\partial E_{q}}n(\beta
E_{q})\right]\delta(E_{{\bf q}+{\bf k}}-E_q-\omega),
\end{eqnarray}
\begin{eqnarray}\label{ImPi_rr}
\Im \Pi_{\rho\rho}(\omega,k)=-\omega\frac{\pi}{A}\sum_{{\bf q}\neq
0}\left[\frac{\hbar^2q^2\nu(q)}{2mE_q}\right]^2
\left[\frac{\partial}{\partial E_{q}}n(\beta
E_{q})\right]\delta(E_{{\bf q}+{\bf k}}-E_q-\omega),
\end{eqnarray}
\begin{eqnarray}\label{RePi_vr}
\Re \Pi_{\varphi\rho}(\omega,k)=\omega\frac{\pi}{A}\sum_{{\bf
q}\neq 0}\frac{\hbar^2 q^2\nu(q)}{2mE_q} \frac{\hbar^2{\bf
kq}}{m}\left[\frac{\partial}{\partial E_{q}}n(\beta
E_{q})\right]\delta(E_{{\bf q}+{\bf k}}-E_q-\omega).
\end{eqnarray}
The above formulae are valid only in the long-wavelength limit.

\end{document}